\documentclass{aa}

\usepackage{graphics}
\usepackage{psfig}
\usepackage{epsfig}

\begin{document}

\title{Gravitational microlensing as a test of stellar model atmospheres}
\titlerunning{Gravitational microlensing as a test of model
atmospheres} 
\author{H.M. Bryce\inst{1,2} 
  \and M.A. Hendry\inst{1} 
  \and D. Valls-Gabaud\inst{3}}
\authorrunning{H.M. Bryce et al.}

\institute{Department of Physics and Astronomy, University of Glasgow,
           Glasgow, G12 8QQ, UK
\and
           203 Van Allen Hall, Department of Physics and Astronomy, 
	   University of Iowa, Iowa City, IA, 52242, USA 
\and
           CNRS UMR 5572, Observatoire Midi-Pyr\'en\'ees, 14 Avenue E. Belin, 
           31400 Toulouse, France}

\offprints{martin@astro.gla.ac.uk}

\date{Received ...; Accepted ...}

\abstract{
We present calculations illustrating the potential of gravitational 
microlensing
to discriminate between classical models of 
stellar surface 
brightness profiles and the recently computed ``Next Generation'' models of
Hauschildt et al. These spherically-symmetric models include a much
improved treatment of molecular lines in the outer atmospheres of cool
giants -- stars which are very 
typical sources
in Galactic bulge microlensing events. We show that the microlensing 
signatures of
intensively monitored point and fold caustic crossing events are readily 
able to distinguish
between NextGen and the classical models, provided a photometric
accuracy of 0.01 magnitudes is reached. This accuracy is now routinely
achieved by alert networks, and hence current observations can
discriminate between such model atmospheres, providing a unique
insight on stellar photospheres.
}

\maketitle

\section{Introduction}
Recently gravitational microlensing has been shown to be an effective
tool for stellar astrophysics, primarily through high time resolution
observations of extended source events -- i.e. where the lensed star
cannot reasonably be approximated as a point source.

The microlensing signatures of extended sources were initially
discussed in Bontz (1979), Gould (1994), Witt \& Mao (1994) and Nemiroff \&
Wickramasinghe (1994) for the case of a star of uniform surface
brightness, where the finite size of the 
star produces a significant deviation from the point source light curve. 
The primary motivation for the authors in the 1990's was to use
extended source effects to better constrain the properties of the
lens -- thus distinguishing between e.g. LMC and galactic lenses.

The microlensing signatures of a non-uniform disc was first considered
by Valls-Gabaud (1994), Bogdanov \& Cherepashchuk (1995), 
Witt (1995) and Simmons, Willis and Newsam
(1995), who showed that the lightcurves would display a chromatic
signature as the lens effectively sees a star of different radius at
different wavelengths.  Valls-Gabaud (1994, 1998) modelled linear,
quadratic and logarithmic limb darkening, with coefficients
computed for the Johnson $U$ to $K$ wavebands using ATLAS 
LTE model atmospheres from Kurucz (1994), and also predicted
a spectroscopic effect, where the line profiles would
change during the microlensing event, providing another probe
of atmospheric structure. This was confirmed by 
Heyrovsk\'y, Sasselov \& Loeb (2000)  for 
a particular red giant model atmosphere. 

Gould (2001) reviews recent observational
progress in detecting non-uniform surface brightness effects. 
The clearest evidence of limb darkening to date
comes from analysis of event MACHO 97-BLG-28 (Albrow et al. 1999). 
This was an example of a binary lens event with a cusp crossing --
arguably the most favourable type of event for stellar
astrophysics studies.
The photometry and sampling coverage for this
event were of sufficient quality to fit a 2-parameter limb darkening
law in $V$ and $I$. 

Although observations already strongly favour limb darkened
atmospheres over uniform brightness models, the use of microlensing as
a discriminant {\em between\/} limb darkened models is a relatively
new subject. Albrow et al. (2001) present a detailed analysis of
event OGLE 99-BUL-23, comparing the $V$ and $I$ light curves with
several different model atmospheres, including for the first
time a treatment of the errors in the lens parameters and their
correlation with the estimated limb darkening
coefficients. Notwithstanding the rigorous statistical approach of
that paper, which `raises the benchmark' for future analyses, only
linear limb darkening models were considered. The current `state of
the art' atmosphere models are the `Next Generation' models of
Hauschildt et al. (1999a, b), which include a much more detailed treatment
of molecular absorption lines in the atmospheres of cool giants and
consider spherically symmetric atmospheres, as opposed to plane
parallel models used previously. The aim of this Letter is to compute 
the microlensing signatures of
typical extended source events, lensed by point and fold caustics.
This allows a comparison of the differing stellar atmosphere models to be
made and so, stimulating interest in observations of extended source
events as a test of model atmospheres.


\section{PHOENIX Next Generation model atmospheres}

Computations of stellar surface brightness profiles have been carried out for
several decades but until very recently were based on  approximate fits
to the limb darkening, that is the dependence of 
 the intensity, $I(\mu)$, as a function of
(the cosine of) emergent angle, $\mu$. For instance 
a simple linear model, namely 
$ I(\mu) = I_0 \left [ 1 - c_1 (1-\mu) \right ] $ 
was often sufficient, while the detailed study of eclipsing binaries
sometimes required  two parameter models, such as the ``square root''
$ I(\mu) = I_0 \left [ 1 - c_2 (1 - \mu) - c_3 ( 1 - \sqrt{\mu} ) \right ] $
or ``logarithmic'' model
$ I(\mu) = I_0 \left [ 1 - c_4 (1 - \mu) - c_5 \mu \ln \mu \right ] $. 
Each coefficient $c_i$  depends on the temperature, gravity
and chemical composition of the source, but the
range of validity of these ``laws'' is often very limited (see
Valls-Gabaud 1998 and references therein). For
instance the square root formulation provides a good fit for
hot stars, while a quadratic equation would be enough for cooler
stars. A new non-linear formulation, using 4 parameters and 
valid across the entire range of effective temperatures and surface gravities
has recently been suggested by Claret (2000). Claret (2000) also
made non-linear fits to the ATLAS  model atmospheres,
improving upon  the previous calculations made by Van Hamme (1993).

By contrast, the recent PHOENIX ``Next Generation'' stellar atmosphere models,
computed by Hauschildt et al. (1999a,b), considerably improve upon these
models
in several important respects, and in particular by carrying out
calculations assuming spherical geometry.
 Moreover, the intensity calculations are based on a huge
library of atomic and molecular 
lines, with about $2 \times 10^8$ molecular lines contributing to a
typical giant atmosphere model at $T_{\rm eff} = 3000 $ K. Claret (2000)
also fitted his new non-linear limb darkening equation to some
PHOENIX models.

The dramatic difference in the dependence of limb darkening on emergent angle 
between the ATLAS models and PHOENIX models is illustrated in
Figure 1, which shows
the intensity profiles for a giant star of $T_{\rm eff} = 4250$ K and
$\log g = 0.5$, in four 
Johnson  bands: $V$, $R$, $I$ and $K$. The solid curve shows the
PHOENIX profiles, while
the dashed, dash-dotted and dotted curves denote the linear,
logarithmic and square root models respectively. It is immediately
clear that there is a dramatic decrease in the predicted intensity 
of the PHOENIX models as one approaches the limb of the star,
i.e. at $\mu \simeq 0.2$. This feature does not appear in ATLAS models and 
arises from the effects of spherical symmetry, which includes finite
optical depth for rays close to the limb -- a treatment which is
absent from ATLAS models  
which utilise plane parallel atmospheres in which the optical depth of
\emph{limb rays} is still finite and so provide significant intensity
out to $\mu =0$. The question 
then arises: is microlensing sufficiently sensitive to 
detect this limb feature in the atmospheres of extended
sources, and thus to test the NextGen models against real observations?

\begin{figure}
\centerline{\psfig{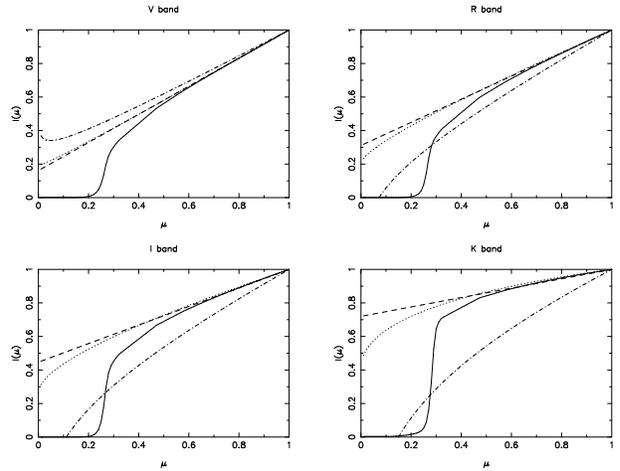}}
\caption[]{\small{Intensity profiles for a giant star of 
$T_{\rm eff} = 4250$ K and $\log g = 0.5$, in four Johnson 
passbands: $V$, $R$,  $I$ and $K$. The solid curve shows the PHOENIX 
profiles, while the dashed, dash-dotted and dotted curves denote the
ATLAS linear, logarithmic and square root models respectively.}}
\label{ng1}
\end{figure}

\section{Computing microlensing lightcurves}
For a point source the amplification due to a point mass lens is
$ A=\left(u^{2}+2\right)/ \left( u\sqrt{u^{2}+4}\right) $, 
where $u(t)$ is the impact parameter, the projected angular 
separation between lens and source,
measured in units of $\theta_E$, the angular Einstein radius of the lens,
defined as
\begin{equation}
\theta_{E} = \sqrt{\frac{4GM}{c^{2}} \frac{(D_{s}-D_{l})}{D_{s}D_{l}}}
\end{equation}
where $D_{l}$ and $D_{s}$ are the lens-observer and source-observer 
separations respectively and $M$ the mass of the lens. If the
impact parameter is comparable to the angular radius of the source, 
the point source amplification function 
breaks down and it becomes necessary to calculate the amplification
as an integral over the source star, given by
\begin{equation}
A_{\nu}(t) =
\frac{\int\int I_{\nu}(r,\theta) \; A(r,\theta,t)\;  r \, dr \, d\theta}
{\int\int I_{\nu}(r,\theta)\; r \, dr \, d\theta}
\label{ng6}
\end{equation} 
Thus, the microlensing lightcurve contains unique information on 
the source 
surface brightness profile, $I_{\nu}(r,\theta)$, which is a far more
sensitive test of the model atmosphere than its integrated value, the
emergent flux. 

In the case of binary lenses, the amplification function takes a more 
complex form, although for an extended source it is again given by an
integral over the source. In this Letter we are only concerned with 
so-called fold caustic crossings, and so we adopt the 
`square-root' approximation for the amplification inside the
caustic structure (Schneider, Ehlers and Falco, 1992) 
which is valid within a few angular source radii of the caustic 
and takes the form
$ A(x) = A_0 + 1/ \sqrt{x} $. 
Here $A_0$ is the total amplification of other images outside the caustic, 
which we assume to be constant during the caustic crossing,
$x$ is the perpendicular distance in units of the angular Einstein
radius of the binary lens, from an element of the source to the
caustic. Note that for elements of the source outside the
caustic structure there are no additional images formed, so $A(x) = A_0$.

\section{Results}
\begin{figure}
\centerline{\psfig{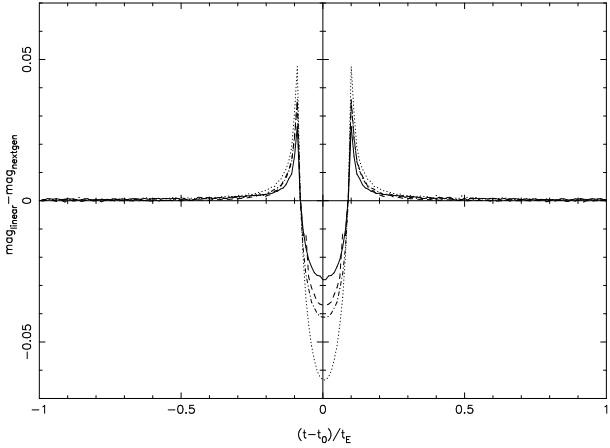}}
\caption[]{\small{A comparison of the $V$, $R$, $I$ and $K$
lightcurves produced by the transit of a point lens with 
minimum impact parameter $u_{0}=0.0$ of a
$4250$ K, $\log g =0.5$, $0.1 \, \theta_{E}$ source, modelled with
PHOENIX and linear limb darkening. Magnitude differences are computed
according to Equation \ref{eq:com} in the text.
}}
\label{ng2}
\end{figure}
\begin{figure}
\centerline{\psfig{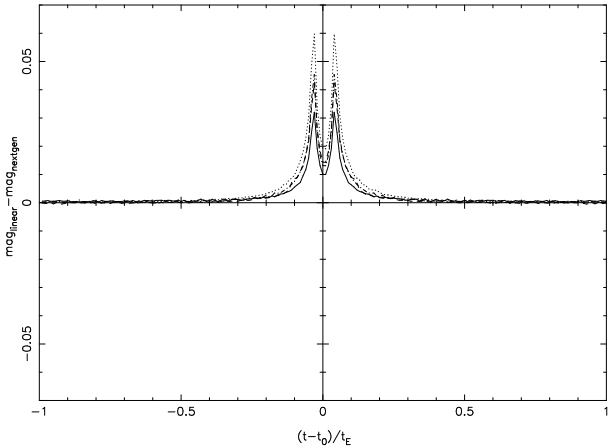}}
\caption[]{\small{A comparison of the $V$, $R$, $I$ and $K$ lightcurves
as in Figure \ref{ng2}, but now for an event with $u_0=0.09$.}}
\label{ng3}
\end{figure}

\begin{figure}
\centerline{\psfig{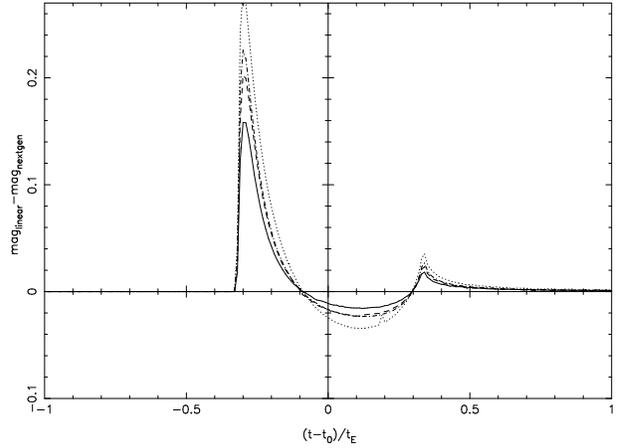}}
\caption[]{\small{A comparison of the $V$, $R$, $I$ and $K$
lightcurves produced by fold caustic crossings of
$4250$ K, $\log g =0.5$, $0.1 \, \theta_{E}$ sources with
PHOENIX and linear limb darkening according to Equation \ref{eq:com}.}}
\label{ng4}
\end{figure}

Figure \ref{ng2} shows a comparison between two sets of microlensing
lightcurves for a point lens transit event: one set calculated using a 
linear limb darkening law and parameters from the ATLAS grid
by Kurucz (1994) (see  Valls-Gabaud, 1998); and 
the second set calculated using the PHOENIX Next Generation models
as discussed in Section 2.  In this example we compare the models for a
source with of angular radius $0.1 \theta_{E}$, $\log g=0.5$, 
$T_{\rm eff} = 4250$ K and solar metallicity. 
The abscissa shows the time evolution of the event, 
in units of the Einstein crossing time, $t_E$, of the lens -- i.e. the time
taken for the lens to move 1 angular Einstein radius. $t_0$ denotes the time
at which the impact parameter is a minimum, $u_0$; in this example we take
$u_0 = 0$ -- i.e. the lens transits through the centre of the source. 
The ordinate shows the magnitude difference between these models, defined as
\begin{equation}
\rm{mag}_{\rm{next gen}}-\rm{mag}_{\rm{linear}}=
2.5 \log \left(\frac{F_{\rm{LL}}}{F_{\rm{UL}}}\right)-2.5\log
\left(\frac{F_{\rm{LN}}}{F_{\rm{UN}}}\right)
\label{eq:com} 
\end{equation}
where $F_{\rm{LL}}$, 
$F_{\rm{UL}}$, $F_{\rm{LN}}$ and $F_{\rm{UN}}$ are the lensed and unlensed
linear limb darkened, and lensed and unlensed NextGen fluxes 
respectively. 
Four colour bands are used to illustrate the strong
chromatic differences between the models: $V$, $R$, $I$ and $K$ bands,
represented by straight, dashed, dash-dotted and dotted lines
respectively.

Two main features are immediately apparent in Figure \ref{ng2}. Firstly, 
just before and after the lens transits the source there are positive
``spikes''; these occur as the PHOENIX/NextGen model is considerably
more limb darkened than the linear model close to the limb. However 
these spikes are very narrow. The second
feature is the broad `brightening' during the central phase of the transit; 
this also occurs as a
consequence of the strong NextGen limb darkening, which 
makes the source appear smaller as the lens crosses its centre, causing a 
larger amplification.  
Thus, if a source with a NextGen type atmosphere is fitted by a
linear limb darkening model, the source size 
would be systematically underestimated.

Figure \ref{ng3} compares lightcurves for the same source as 
Figure \ref{ng2} but with an impact 
parameter of $0.09 \, \theta_{E}$, i.e. the lens `just' transits
the source.   
In this case the only features are the two upward spikes, due to the NextGen 
intensity being considerably smaller than the linear model intensity
at the limb of the  star. Again, however, these spikes are very narrow
and around  minimum impact parameter the difference between the models is 
much smaller, since the amplified flux is dominated by a 
region slightly closer to the centre of the source where there is less 
difference between the models (see Figure \ref{ng1} at $\mu = 0.9$). In
particular, no brightening effect is seen since the lens does 
not cross the central region of the source. The amplitude of the
effect is, however, similar: at least $0.01$ mag, which is easily detectable.

Figure \ref{ng4} compares the NextGen and linear models during a fold
caustic crossing event, shown from when the source is 
$3$ source radii outside to $3$ source
radii inside the caustic structure. 
Here the source has 
$T_{\rm eff} = 4250$ K, $\log g = 0.5$ and solar metallicity, 
but now has an angular radius of 
$0.01 \, \theta_{E}$ -- a more realistic value for e.g. typical Bulge 
events, which results in larger amplification changes.  The most striking
feature on Figure \ref{ng4} is the large spike as the source begins to enter 
the caustic; here the flux is
dominated by the small region of photosphere very close to 
the limb and directly beneath the caustic -- the surface brightness of
which varies dramatically between the two models. As the event 
proceeds the magnitude difference becomes much smaller, particularly 
in mid-transit when the limb differences 
between the models are diluted by the larger intensity from the central part of
the source. However, we again see a broad `brightening' effect as in
Figure \ref{ng2}, due to the extreme NextGen limb darkening 
producing an apparently smaller source.  As 
the trailing limb crosses the caustic we see only a small peak 
since there is already considerable amplification
from the rest of the photosphere. In practice, of course, we would expect to
observe these events in reverse: it is easier to predict
when a source will exit a caustic than when it will enter!

\section{Conclusions}

Figures 2 -- 4 clearly suggest that it is within the capabilities of current
intensive microlensing monitoring programs to discriminate between the
chosen atmosphere models from a well parameterised lightcurve. For the
case of a point caustic
event, with e.g. timescale $t_{E} \simeq 15$ days, the
transits shown in Figure
2 and 3 would last for about 3 days, with the sharp features
at each limb crossing a few hours in duration. The
sampling rates regularly achieved
by e.g. the PLANET collaboration during recent events would easily suffice to
resolve these limb crossing features with a reasonable number of data
points. It must be noted, of course, that since Figures 2 and 3 are
for point lens transits, the probability
of such an event -- and indeed of its immediate detection -- is
small. \emph{Given} that the event is alerted during the rise phase
as a potential extended source, however, the
detection of the limb crossing spikes would be highly likely. 

The situation for the fold caustic event is more encouraging. Again,
the timescale for
the caustic crossing is of the order of several hours, which is
typical of ``lensing anomaly''
events already regularly monitored. Moreover, as pointed out
previously, the chief advantage
of fold caustic crossing events is the predictability of the source
exiting the caustic structure,
thus allowing intensive monitoring to be scheduled in advance. As the
trailing limb of the
star leaves the interior of the caustic one will see a dramatic
difference between  NextGen and the
other models since it is effectively \emph{only} the (strongly
darkened) limb of the star which
is being amplified. The signature of this limb crossing
is seen in Figure 4 to be
of order 0.2 mag. in Johnson $B$, $V$, $R$ and $I$ -- which would 
render it very easily detectable with existing photometric
precision.

To compare any stellar atmosphere models using
real data would involve
simultaneously determining best-fit lens parameters for the event
under study. As remarked previously,
Albrow et al (2001) have included the effects of lens parameter
errors in estimating limb darkening
coefficients. We will carry out a similar numerical study, using
realistic simulated observations,
in future work and at a more advanced stage one must also consider the
possibility of spots further effecting the lightcurve (see Hendry et
al. 2002).  It seems clear, however, that the magnitude of the
difference between NextGen and ATLAS models is eminently
detectable -- even allowing for
uncertainties in the fitted lens parameters. This is an essential
step in the determination of fundamental stellar parameters.
For instance, high accuracy measures of stellar radii via 
eclipsing binary light curves rely critically on these
model atmospheres (e.g. Orosz and Hauschildt 2000) as do
interferometric measures (e.g. Davis et al. 2000, Wittkowski et al. 2001).
Thanks
to microlensing we have now almost direct access to the distribution
of intensity across a source, and hence we can probe
stellar photospheres with unprecedented detail.

\end{document}